\documentclass[12pt]{article}
\usepackage{graphicx,amssymb,amsmath,empheq,amsthm}
\usepackage{authblk}
\usepackage{comment}
\usepackage{braket}
\usepackage{subfig} 
\usepackage{soul,}
\usepackage{hyperref}
\hypersetup{colorlinks = true, linkcolor=black, citecolor=black, urlcolor=blue}
\usepackage{url}
\usepackage{enumitem}
\usepackage[nosort]{cite}
\usepackage{amsmath,bm}

\theoremstyle{plain}

\usepackage{tikz}
\usepackage{tikz-cd}
\usetikzlibrary{arrows}
\usetikzlibrary{intersections}
\usetikzlibrary{shapes.geometric}
\usetikzlibrary{decorations.pathmorphing, patterns,shapes}
\usetikzlibrary{decorations.markings}

\tikzset{
	mid arrow/.style={postaction={decorate,decoration={
				markings,
				mark=at position .575 with {\arrow{stealth}}
	}}},
	near arrow/.style={postaction={decorate,decoration={
				markings,
				mark=at position .275 with {\arrow{stealth}}
	}}},
	far arrow/.style={postaction={decorate,decoration={
				markings,
				mark=at position .800 with {\arrow{stealth}}
	}}},
	snake arrow/.style={fixed point arithmetic, decorate, decoration={snake,amplitude=2pt, segment length=11pt},postaction={decoration={markings,mark=at position 0.625 with {\arrow{stealth}}},decorate}},
}

\tikzset{
  baseline = -0.5ex,
  wavy/.style = {
    thick,
    decorate,
    decoration={snake,amplitude=2pt,segment length=5pt}},
  sdot/.style = {
    circle,
    draw=none,
    fill=black,
    minimum size=2.5pt,
    inner sep=0pt},
  bdot/.style = {
    circle,
    draw=none,
    fill=black,
    minimum size=4pt,
    inner sep=0pt},
  svertex/.style = {
    circle,
    draw=black,
    thick,
    fill=lightgray,
    minimum size=8pt,
    inner sep=1pt},
  bvertex/.style = {
    circle,
    draw=black,
    thick,
    fill=lightgray,
    minimum size=16pt},
  dvertex/.style = {
    circle,
    draw=black,
    thick,
    fill=gray,
    minimum size=25pt}}

\title{Hybrid Brownian SYK–Hubbard Model: from Spectral Function to Quantum Chaos}

\author[1]{Ning Sun}
\author[2]{Peng Zhang\thanks{pengzhang@ruc.edu.cn}}
\author[1,3]{Pengfei Zhang\thanks{pengfeizhang.physics@gmail.com}}

\affil[1]{\normalsize \it Department of Physics \& State Key Laboratory of Surface Physics, Fudan University, Shanghai, 200438, China}
\affil[2]{\normalsize \it School of Physics \& Key Laboratory of Quantum State Construction and Manipulation (Ministry of Education), Renmin University of China, Beijing, 100872, China}
\affil[3]{\normalsize \it Hefei National Laboratory, Hefei 230088, China}

\date{\today}

\begin{document}
\maketitle

\begin{abstract}
Understanding the emergence of complex correlations in strongly interacting systems remains a fundamental challenge in quantum many-body physics. One fruitful approach is to develop solvable toy models that encapsulate universal properties shared by realistic systems. In this work, we introduce the Brownian SYK-Hubbard model, which combines the all-to-all random interactions of the Sachdev–Ye–Kitaev (SYK) model with on-site Hubbard-type interactions. This hybrid construction enables the study of the interplay between nonlocal random dynamics and local correlation effects: (1) As the interaction strength increases, the single-particle spectrum exhibits a transition from a single peak to a two-peak structure, signaling the onset of Mottness. (2) The spectral form factor undergoes a sequence of dynamical transitions as the evolution time increases before reaching the plateau in the long-time limit under strong Hubbard interactions. (3) The out-of-time-order correlator is computed by summing a series of modified ladder diagrams, which determines the quantum Lyapunov exponent and reveals a violation of the bound on branching time. Our results establish a new analytically tractable platform for exploring the effects of Hubbard interactions in chaotic many-body systems.
\end{abstract}

\bigskip

\section{Introduction}
The exponential growth of the Hilbert space dimension with increasing system size poses a fundamental challenge for understanding quantum many-body systems, restricting numerical simulations to moderate sizes. An alternative approach is to study solvable toy models, whose physical properties can be analyzed using algebraic or field-theoretical methods. A prominent example is the Sachdev–Ye–Kitaev (SYK) model, which describes $N$ Majorana fermions with random $q$-body interactions \cite{sachdev1993gapless,kitaev2015simple,Maldacena:2016hyu}. It has attracted extensive attention across high-energy physics, condensed matter physics, and quantum information. Using the large-$N$ expansion, the system's correlation functions can be expressed through self-consistent equations, revealing the emergence of non-Fermi liquid behavior \cite{Maldacena:2016hyu,PhysRevB.95.134302,PhysRevLett.119.216601,PhysRevLett.119.207603,RevModPhys.94.035004} and maximal chaos \cite{Maldacena:2015waa} in the low-temperature limit for $q \geq 4$. This maximal chaos also suggests the possibility of a holographic description of the low-energy sector, as established in Ref. \cite{Maldacena:2016upp}. Subsequent studies have extended the analysis beyond correlation functions to quantities such as the spectral form factor (SFF) \cite{Saad:2018bqo} and quantum entanglement \cite{Huang:2017nox,Penington:2019kki,Dadras:2020xfl,Chen:2020wiq,Zhang:2020kia,Zhang:2022yaw}. Beyond the original SYK model, various extensions have been proposed to explore different physical phenomena. For instance, in the Brownian SYK model \cite{Saad:2018bqo,Sunderhauf:2019djv}, the static random interactions are replaced by time-dependent Brownian interactions, which further allow the analytical determination of correlation functions even at high temperatures \cite{Zhang:2020jhn}.

Most generalizations of the SYK model require fully random interactions, as in the traditional SYK model, which are compatible with the standard tensor large-$N$ structure \cite{Witten:2016iux}. On the other hand, introducing interactions without randomness can be fruitful from several complementary perspectives. First, the traditional ladder-diagram structure of the out-of-time-order correlator imposes a fundamental bound on the branching time \cite{Gu:2018jsv,Zhang:2020jhn}, which serves as an obstacle to achieving sub-AdS holography. Breaking this ladder-diagram structure necessitates the inclusion of constant (non-random) interactions. Second, many intriguing physical phenomena are inherently associated with constant interactions. A prominent example is the emergence of the Mott insulating phase, which originates from the Hubbard interaction \cite{RevModPhys.70.1039,RevModPhys.78.17}. Given the fundamental interest in the interplay between Hubbard interactions and superconductivity, incorporating such interactions into concrete solvable models is particularly important for developing a deeper understanding of realistic materials. Third, designing models that combine both random and constant interactions can help identify broader classes of solvable models with novel underlying structures. 

\begin{figure}[t]
    \centering
    \includegraphics[width=0.55\linewidth]{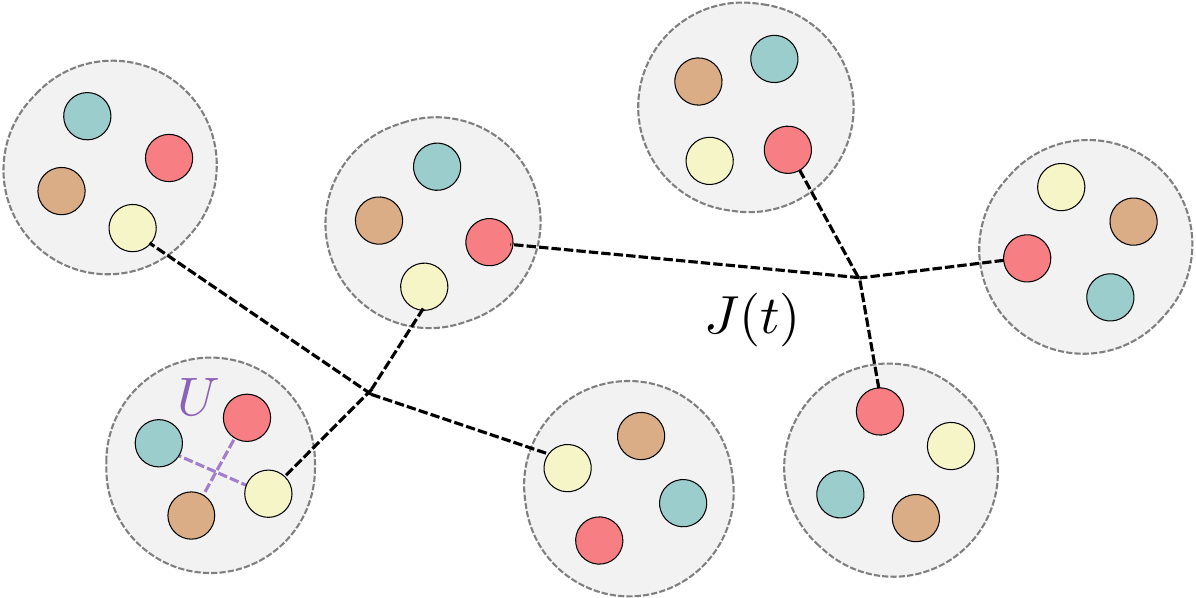}
    \caption{ We present a schematic illustration of the Brownian SYK–Hubbard model studied in this work. The model consists of $4N$ Majorana fermions grouped into $N$ sites, with each site hosting four distinct modes labeled by different colors. Modes sharing the same color interact through Brownian SYK-type random couplings $J^a_{i_1,...,i_q}(t)$, while modes residing on the same site are coupled via a constant on-site Hubbard interaction $U$. }
    \label{fig:schematic}
  \end{figure}

With this motivation, we introduce the Brownian SYK–Hubbard model as a novel platform for investigating the interplay between Brownian SYK-type random couplings and constant Hubbard interactions (see Figure \ref{fig:schematic} for an illustration). This hybrid model admits analytical treatment for a broad range of physical observables defined on Keldysh contours with a small number of replicas \cite{kamenev2023field,Aleiner:2016eni}. For single-replica quantities, we compute the two-point function and single-particle spectrum, revealing a transition from a single-peak to a double-peak structure. The emergence of the double-peak spectrum serves as a hallmark of Mottness familiar from correlated solid-state systems. We further analyze the spectral form factor (SFF), which exhibits a sequence of dynamical transitions at longer evolution times induced by strong Hubbard interactions. For two-replica observables, we evaluate the out-of-time-order correlator (OTOC) by identifying a family of generalized ladder diagrams that fully capture the effects of the Hubbard interaction. We explicitly demonstrate that the model violates the conventional bound on the branching time characteristic of SYK-like systems. Altogether, our results open a new avenue for exploring the rich consequences of incorporating Hubbard interactions into analytically tractable, strongly interacting many-body models.

\section{Model and Two-point Function}

The original SYK model describes $N$ Majorana modes with random interactions and exhibits full permutation symmetry after disorder averaging. To hybridize SYK interactions with Hubbard interactions while preserving analytical solvability, we instead consider a system of $N$ sites labeled by $i \in \{1, 2, \ldots, N\}$. Each site hosts four distinct Majorana fermion modes $\chi_{ia}$, labeled by $a \in \{1, 2, 3, 4\}$, which are illustrated as different colors in Figure \ref{fig:schematic}. We impose the canonical anticommutation relation $\{\chi_{ia}, \chi_{jb}\} = \delta_{ij}\delta_{ab}$. The Hamiltonian reads
\begin{equation}\label{eq:H}
H(t)=i^{q(q-1)/2}\sum_a\sum_{i_1<i_2<...< i_q}J_{i_1,i_2,...,i_q}^a(t)\chi_{i_1a}\chi_{i_2a}...\chi_{i_qa}+\sum_{i}U\chi_{i1}\chi_{i2}\chi_{i3}\chi_{i4}.
\end{equation}
The first term represents the standard Brownian SYK interaction for each flavor $a$, where the factor $i^{q(q-1)/2}$ ensures the Hermiticity of the Hamiltonian. The random Brownian couplings $J_{i_1,i_2,...,i_q}^a(t)$ satisfy
\begin{equation}
\overline{J_{i_1,i_2,...,i_q}^a(t)}=0,\ \ \ \ \  \overline{J_{i_1,i_2,...,i_q}^a(t)^2}=\frac{(q-1)!J}{N^{q-1}dt}.
\end{equation}
The second term in \eqref{eq:H} corresponds to the Hubbard interaction expressed in the Majorana fermion representation, which exhibits an $SO(4)$ symmetry among the four Majorana modes \cite{1990MPLB....4..759Y}. More explicitly, if we introduce complex fermions $c_{j,\uparrow}=(\chi_{j1}+i\chi_{j2})/\sqrt{2}$ and $c_{j,\downarrow}=(\chi_{j4}+i\chi_{j3})/\sqrt{2}$, we find 
\begin{equation}\label{eq:Hubbardint_explicit}
U\chi_{i1}\chi_{i2}\chi_{i3}\chi_{i4}=U\left(c_{i,\uparrow}^\dagger c_{i,\uparrow}-\frac{1}{2}\right)\left(c_{i,\downarrow}^\dagger c_{i,\downarrow}-\frac{1}{2}\right).
\end{equation}
This is the standard on-site Hubbard interaction at site $i$, with the chemical potential tuned to half-filling, $\mu = U/2$. 

Before turning to the explicit analysis, let us first clarify the intuition behind the solvability of this model. Both the traditional SYK model and its Brownian variant can be systematically analyzed using the large-$N$ expansion. For example, the self-energy of the two-point function is organized into a series of melonic diagrams. This structure, however, is disrupted by the presence of the Hubbard interaction, which generates a variety of additional diagrams contributing to the self-energy. Instead, one may adopt a different perspective \cite{Kitaev:2017awl}: the solvability of the SYK model originates from approximating the effect of random couplings between different sites as that of a single mode embedded in an effective environment, where the environment operator $$\xi_i^a(t)\approx \sum_{i_2<...< i_q}J_{i,i_2,...,i_q}^a(t)\chi_{i_2a}...\chi_{i_qa}$$ captures the averaged influence of all other sites. In static SYK models, the correlation function of $\xi_i^a(t)$, which characterizes the environment, generally depends on the dynamics of each mode. In contrast, in the Brownian SYK case, this correlation function reduces to a simple delta function in time, reflecting its Brownian nature. After introducing the Hubbard interaction, a similar analysis can be carried out by retaining four modes within a single site, each coupled to the effective environment. This reformulation reduces the original many-body problem to that of an open quantum system with a finite number of modes. Even without further simplification, the effective single-site problem can be directly simulated using exact diagonalization.

We now present our calculation of the two-point function. Since the Hamiltonian is time-dependent, the system is always in an effective infinite-temperature state. Consequently, the two-point function of interest is
\begin{equation}
G(t)=\overline{\langle \chi_{ia}(t)\chi_{ia}(0)\rangle}=2^{-2N}\overline{\text{tr}[U^\dagger (t)\chi_{ia}U(t)\chi_{ia}]}.
\end{equation}
where the time-evolution operator is defined as $U(t) = \mathcal{T} \exp\Big(-i \int_0^t H(t'), dt'\Big)$, with $\mathcal{T}$ denoting the time-ordering operator. The Green’s function is independent of both $i$ and $a$ owing to the permutation symmetry of the system after disorder averaging. To compute the Green's function, we consider the Keldysh contour represntation for the partition function $1=\overline{Z}=\overline{\text{tr}[U(t)U^\dagger (t)]}$. We have
\begin{equation}\label{eq:single_K}
\begin{aligned}
Z&=\ \ \begin{tikzpicture}[scale=1.3, baseline={([yshift=-2pt]current bounding box.center)}]
\draw[thick] (20pt,5pt) arc(90:-90:5pt and 5pt);
\draw[thick] (-20pt,5pt) arc(90:270:5pt and 5pt);
\draw[mid arrow,thick]  (-20pt,5pt)  --  (20pt,5pt);
\draw[mid arrow,thick]  (20pt,-5pt)  --  (-20pt,-5pt);
\filldraw (25pt,0pt) circle (1pt);
\end{tikzpicture}=\int \mathcal{D}\chi_{ia}^s \exp\left(-\sum_{sia}\int dt~s\frac{1}{2}\chi^s_{ia}\partial_t\chi^{s}_{ia}-i\sum_s\int dt~sH[\chi_{ia}^s(t)]\right).
\end{aligned}
\end{equation}
Here, $s = \pm$ labels the branches corresponding to forward and backward time evolution. We fix the contour ordering by taking the initial point for the path-integral to be the black dot. Next, we perform the disorder average over the Brownian variables and introduce the auxiliary equal-time self-energy fields $\Sigma_a^{ss'}(t)$ and Green’s function fields $G_a^{ss'}(t)$, following the standard SYK methodology \cite{Maldacena:2016hyu}. This finally leads to the action
\begin{equation}\label{eqn:action_single}
S=\int dt~\left[\frac{1}{2}\chi^s_{ia}(s\delta_{ss'}\partial_t-\Sigma_{a}^{ss'})\chi^{s'}_{ia}+\frac{1}{2}\Sigma_{a}^{ss'}G_{a}^{ss'}+\frac{Jss'}{2q}(G_{a}^{ss'})^q+iUs\chi^s_{i1}\chi^s_{i2}\chi^s_{i3}\chi^s_{i4}\right].
\end{equation}
Here, we keep the summation over indices implicit. The partition function is given by $\overline{Z}=\int  \mathcal{D}\chi_{ia}^s  \mathcal{D}\Sigma_a^{ss'}\mathcal{D}G_a^{ss'}~\exp(-S)$. Unlike the traditional SYK model, this integration cannot be carried out for a general value of $U$. Consequently, there is no explicit expression for the $G$–$\Sigma$ action in the Brownian SYK–Hubbard model. Nevertheless, we can still perform a saddle-point approximation for both $\Sigma_a^{ss'}(t)$ and $G_a^{ss'}(t)$. This is because both fields are independent of the large-$N$ site index $i$, so that their effective action, although not explicitly known, contains an overall factor of $N$. As a result, in the limit $N \rightarrow \infty$, the saddle-point contribution dominates. Performing the variation with respect to $\Sigma_a^{ss'}(t)$ and $G_a^{ss'}(t)$, we obtain the saddle-point equations
\begin{equation}
G_a^{ss'}(t)=\langle \chi^s_{ia}(t) \chi^{s'}_{ia}(t)\rangle,\ \ \ \ \ \ \Sigma_a^{ss'}(t)=JG_a^{ss'}(t)^{q-1}.
\end{equation}
Using $(\chi_{ia})^2 = 1/2$ (follows from the canonical anticommutation relation) and the unitarity of time evolution, we obtain explicit solutions for both equations as
\begin{equation}\label{eq:sol_singleK}
\Sigma_{a}^{+-}(t)=-\Sigma_{a}^{-+}(t)=\frac{J}{2^{q-1}}\equiv\frac{\Gamma_0}{2},\ \ \ \ \ \ G_{a}^{+-}(t)=-G_{a}^{-+}(t)=\frac{1}{2}.
\end{equation}
Here, $\Gamma_0$ denotes the decay rate of the Brownian SYK model \cite{Zhang:2020jhn}, while all other components of $\Sigma_a^{ss'}(t)$ and $G_a^{ss'}(t)$ vanish. Consequently, to leading order in $1/N$, we can fix all auxiliary fields at their saddle-point values and obtain an effective action for fermions
\begin{equation}\label{eq:single_Seff}
S_{\text{eff}}=\sum_i\int dt~\left[\sum_{a,ss'}\frac{1}{2}\chi^s_{ia}\left(s\delta_{ss'}\partial_t-i\frac{\Gamma_0}{2}(\sigma_y)_{ss'}\right)\chi^{s'}_{ia}+iU\sum_ss\chi^s_{i1}\chi^s_{i2}\chi^s_{i3}\chi^s_{i4}\right]+\text{cons.}
\end{equation}
The partition function reads $\overline{Z}=\int  \mathcal{D}\chi_{ia}^s~\exp(-S_{\text{eff}})$. 

We can make two important observations. First, modes corresponding to different sites (i.e., different values of $i$) decouple in the effective action. Therefore, we can focus on eight fields ($a \in \{1, 2, 3, 4\}$ and $s = \pm$) associated with a single site. Second, the evolution of a single site coincides with that described by a Lindblad master equation with jump operators $\chi_{ia}$ \cite{PhysRevResearch.4.L022068,PhysRevD.107.106006,PhysRevB.106.075138,PhysRevD.109.046005}. This correspondence can be made explicit through the operator–state mapping \cite{Qi:2018bje,Gu:2017njx}, which leads to an expectation under the non-unitary evolution
\begin{equation}\label{eq:partitionfunc_single_replica}
\begin{aligned}
\overline{Z}&=\begin{tikzpicture}[scale=1.3, baseline={([yshift=-2pt]current bounding box.center)}]
\draw[thick] (20pt,5pt) arc(90:-90:5pt and 5pt);
\draw[thick] (-20pt,5pt) arc(90:270:5pt and 5pt);
\draw[dotted,thick] (-18pt,7pt) -- (-18pt,-7pt);
\draw[dotted,thick] (18pt,7pt) -- (18pt,-7pt);
\draw[mid arrow,thick]  (-20pt,5pt)  --  (20pt,5pt);
\draw[mid arrow,thick]  (-20pt,-5pt)  --  (20pt,-5pt);
\end{tikzpicture}~\propto\bra{\text{EPR}}e^{H^{(1)}_{\text{eff}}t}\ket{\text{EPR}}^N,\\
H^{(1)}_{\text{eff}}&=i\sum_a\frac{\Gamma_0}{2}\chi_{a}^L\chi_{a}^R-iU\chi_{1}^L\chi_{2}^L\chi_{3}^L\chi_{4}^L+iU\chi_{1}^R\chi_{2}^R\chi_{3}^R\chi_{4}^R.
\end{aligned}
\end{equation}
Here, we introduce two copies of the Majorana fermion system for a representative site $j_0$ with Majorana operators $\chi^{L/R}_a$. Loosely speaking, we could identify the operator $(\chi_{a}^L,i\chi_{a}^R)$ with the field $(\chi_{j_0a}^+,\chi_{j_0a}^-)$. The EPR state is defined by the condition $(\chi_{a}^L-i\chi_{a}^R)\ket{\text{EPR}}=0$. It is straightforward to verify that this EPR state is an eigenstate of the non-Hermitian effective Hamiltonian $H^{(1)}_{\text{eff}}$ with the maximal real eigenvalue $2\Gamma_0$. Moreover, since the model contains only eight Majorana fermions, the full spectrum of $H^{(1)}_{\text{eff}}$ can be readily obtained via exact diagonalization.

  \begin{figure}[t]
    \centering
    \includegraphics[width=0.83\linewidth]{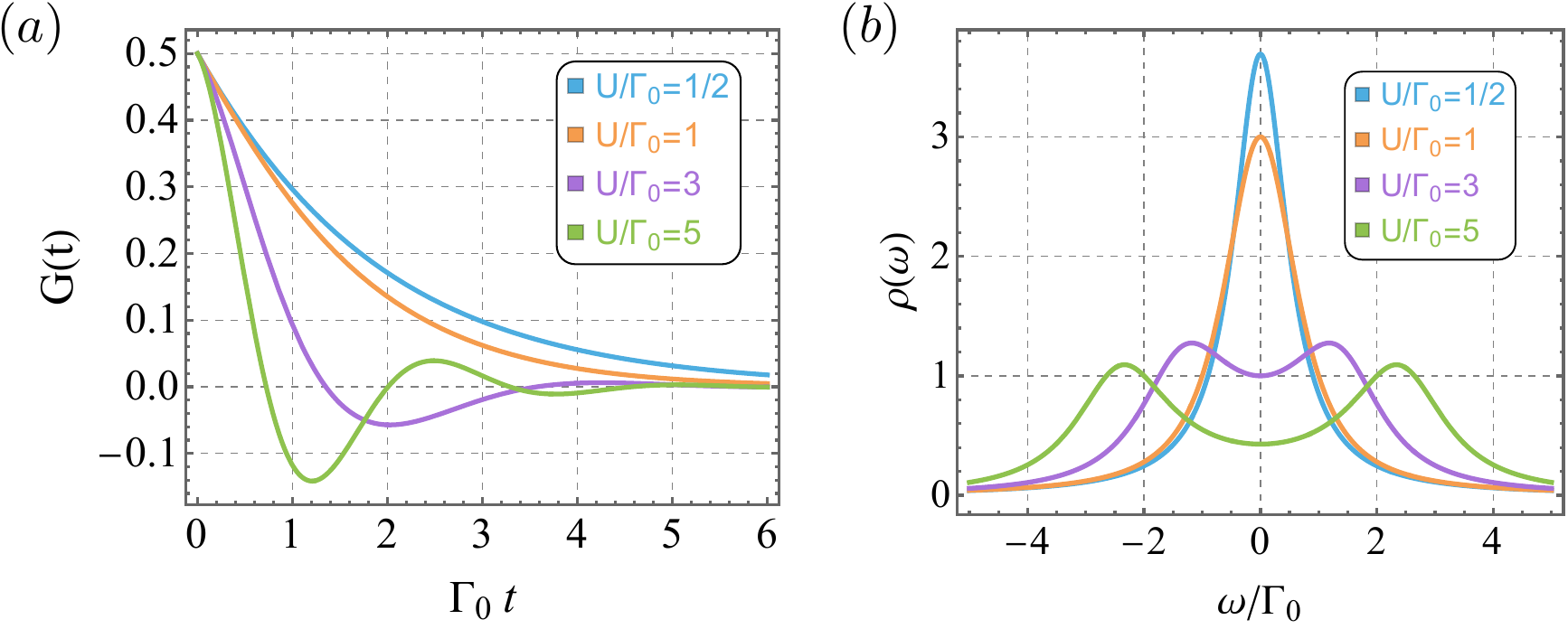}
    \caption{ We plot the Green’s function and spectral function of the Brownian SYK–Hubbard model (with arbitrary $q$) for $U/\Gamma_0 \in \{1/2, 1, 3, 5\}$. The results clearly show a qualitative change at $U/\Gamma_0 = 1$, where the Green’s function transitions from monotonic decay to oscillatory behavior, and the spectral function evolves from a single peak to double peaks.}
    \label{fig:two-point}
  \end{figure}

This allows us to compute the exact two-point function $G(t)$ in the large-$N$ limit. The Green's function involves additional insertions of Majorana fields in the path integral, which modifies equation \eqref{eq:partitionfunc_single_replica} as
\begin{equation}\label{eq:resG}
\begin{aligned}
G(t)&=\overline{\langle \chi_{ia}(t)\chi_{ia}(0)\rangle}=\frac{\bra{\text{EPR}}\chi_1^Le^{H^{(1)}_{\text{eff}}t}\chi_1^L\ket{\text{EPR}}}{\bra{\text{EPR}}e^{H^{(1)}_{\text{eff}}t}\ket{\text{EPR}}}\\&=\frac{1}{2} e^{-\Gamma_0  t} \left(\frac{\Gamma_0  \sinh \left(\frac{\sqrt{\Gamma_0 ^2-U^2}}{2} t \right)}{\sqrt{\Gamma_0 ^2-U^2}}+\cosh \left(\frac{\sqrt{\Gamma_0^2-U^2}}{2} t \right)\right).
   \end{aligned}
\end{equation}
Here, the second line is obtained via direct exact diagonalization. For $U = 0$, the result reduces to $G(t) = \frac{1}{2} e^{-\Gamma_0 t / 2}$, in agreement with the known result for the Brownian SYK model \cite{Zhang:2020jhn}. We plot the Green’s function for several different $U/\Gamma_0$ in Figure \ref{fig:two-point}(a). As $U$ increases, the decay rate at large $t$, defined via $G(t) \sim e^{-\Gamma t/2}$, initially decreases following the expression $\Gamma = 2\Gamma_0 - \sqrt{\Gamma_0^2 - U^2}$, reaching the maximal value $\Gamma = 2\Gamma_0$ at $U = \Gamma_0$. For $U > \Gamma_0$, the exponent becomes complex, and the Green’s function exhibits oscillations with frequency $\omega = \frac{1}{2} \sqrt{U^2 - \Gamma_0^2}$. Meanwhile, the decay rate remains fixed at $\Gamma = 2\Gamma_0$. Similar transitions between monotonic decay and oscillatory Green’s functions have also been observed both theoretically and experimentally in random spin models \cite{Li:2024pta,Zhang:2023wtr,Tian-GangZhou:2023lzb}. In the large-$U$ limit, the oscillation frequency approaches $U/2$, corresponding to the excitation energy of the Hubbard term \eqref{eq:Hubbardint_explicit}. 

The dynamical transition also manifests in the single-particle spectral function, which is related to the Green’s function via a Fourier transform. Using \eqref{eq:resG}, we obtain the exact result
\begin{equation}
\rho(\omega)=\frac{4 \Gamma _0 \left(9 \Gamma _0^2+3 U^2+4 \omega ^2\right)}{9 \Gamma _0^4+\Gamma _0^2
   \left(6 U^2+40 \omega ^2\right)+\left(U^2-4 \omega ^2\right)^2}.
\end{equation} 
The spectral function is plotted in Fig.~\ref{fig:two-point}(b). It exhibits a single peak at $\omega=0$ for $U\leq \Gamma_0$, whose width increases with $U$. For $U> \Gamma_0$, two peaks emerge at finite $\omega$. This double-peak structure is analogous to that of a Mott insulator’s spectral function. Nevertheless, the Brownian SYK interaction fills the Mott gap, and the system remains gapless for any $U/\Gamma_0$.

Finally, we note that our calculation explicitly demonstrates that the two-point function of the Brownian SYK–Hubbard model exactly coincides with that of a four-Majorana Lindblad evolution. However, this equivalence holds only for the Keldysh contour \eqref{eq:single_K}, which corresponds to the simple saddle-point solution \eqref{eq:sol_singleK}. For more complex physical observables, the intrinsic chaotic nature of the Brownian SYK dynamics is expected to play a crucial role, as we will discuss in later sections.

\section{Spectral Form Factor}
For a chaotic system with a static Hamiltonian $H$, the spectral form factor, defined as $\text{SFF}(T) \equiv |\mathrm{Tr}(e^{-iHT})|^2$, has been introduced to capture the universal random-matrix statistics of energy levels \cite{haake1991quantum,Cotler:2016fpe}. As time increases, the SFF initially decays in a non-universal “slope” region, which becomes self-averaging in disordered models. In the SYK model, this early-time decay originates from fluctuations around the naive saddle point, where the forward and backward branches of time evolution remain disconnected \cite{Cotler:2016fpe}. At later times, a new saddle point emerges in which these two branches become connected, leading to the so-called “ramp” regime, where the SFF grows linearly due to the increasing number of connected saddles. The “ramp” regime reflects the universal level repulsion that underlies the spectral correlations of chaotic Hamiltonians. Finally, at very late times, the SFF saturates into a “plateau” regime, whose behavior arises from highly non-perturbative contributions that requires sophisticated resummation \cite{Saad:2022kfe}. 

For systems with Brownian interactions, the definition of the spectral form factor (SFF) is generalized as $\text{SFF}(T) \equiv |\mathrm{Tr}(U(T))|^2$. Unlike systems with static Hamiltonians, the “slope” regime is directly followed by the “plateau” regime. This latter regime, analogous to the “ramp” regime in static Hamiltonians, can be understood in terms of the emergence of new saddle points. These features are explicitly demonstrated in the Brownian SYK model in Ref. \cite{Cotler:2016fpe}. Here, we explore the consequence of introducing Hubbard interactions. The path-integral presentation of the SFF reads
\begin{equation}\label{eq:defsff}
\overline{\text{SFF}}(T)=\ \begin{tikzpicture}[scale=1.3, baseline={([yshift=-2pt]current bounding box.center)}]
\draw[mid arrow,thick]  (-20pt,5pt)  --  (20pt,5pt);
\draw[mid arrow,thick]  (20pt,-5pt)  --  (-20pt,-5pt);
\draw[thick]  (18pt,-7pt)  --  (18pt,-3pt);
\draw[thick]  (-18pt,-7pt)  --  (-18pt,-3pt);
\draw[thick]  (17pt,7pt)  --  (17pt,3pt);
\draw[thick]  (-17pt,7pt)  --  (-17pt,3pt);
\draw[thick]  (19pt,7pt)  --  (19pt,3pt);
\draw[thick]  (-19pt,7pt)  --  (-19pt,3pt);
\end{tikzpicture}=\int_{\text{PBC}}  \mathcal{D}\chi_{ia}^s  \mathcal{D}\Sigma_a^{ss'}\mathcal{D}G_a^{ss'}~\exp(-S),
\end{equation}
where the action $S$ is again given by \eqref{eqn:action_single}. Here, we use single and double slash lines to denote the periodic boundary conditions for the forward and backward evolution branches, respectively. Following the same reasoning as in the previous section, one can perform a saddle-point analysis analogous to that on the Keldysh contour. However, due to the difference in boundary conditions, the forward evolution no longer cancels the backward evolution. Consequently, the saddle-point values of $\Sigma_a^{+-}$ and $G_a^{+-}$ are not fixed. We therefore introduce the parametrization
\begin{equation}
\Sigma_{a}^{+-}=-\Sigma_{a}^{-+}=\frac{\lambda}{2},\ \ \ \ \ \ G_{a}^{+-}=-G_{a}^{-+}=\frac{g}{2}.
\end{equation}
Here, we assume that the saddle-point solution preserves the time-translation symmetry after the disorder average.

  \begin{figure}[t]
    \centering
    \includegraphics[width=1\linewidth]{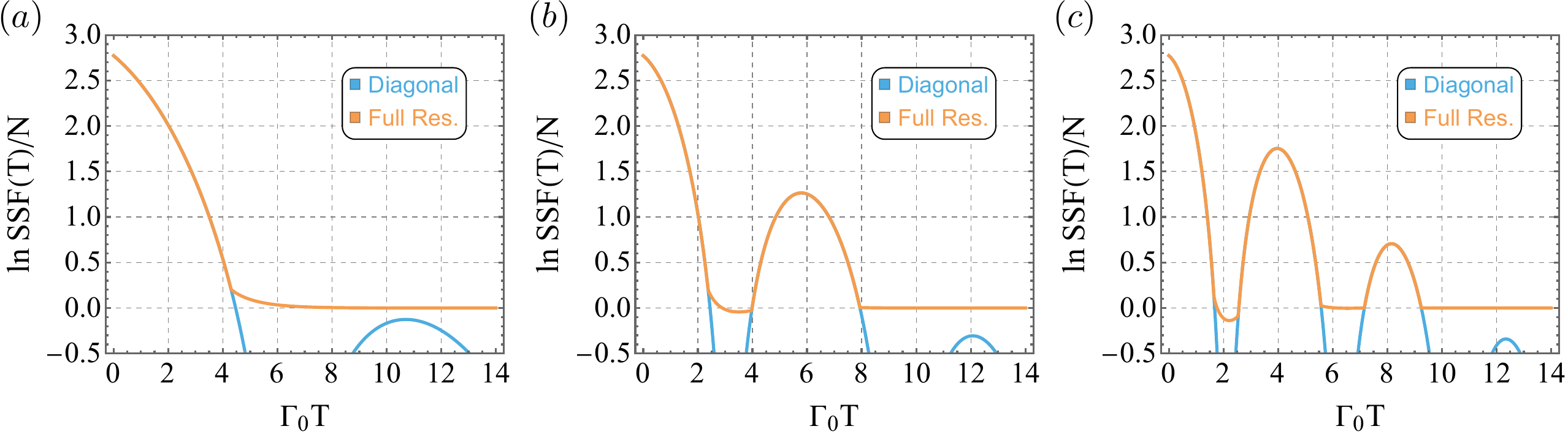}
    \caption{ We plot the SFF of the Brownian SYK–Hubbard model with $q=4$ for (a) $U/\Gamma_0 = 1$, (b) $U/\Gamma_0 = 2$, and (c) $U/\Gamma_0 = 3$. The number of dynamical transitions between the diagonal and connected saddles increases with $U$, due to the persistent oscillations in the diagonal contribution.  }
    \label{fig:SFF}
  \end{figure}

With this parametrization, the SFF can again be expressed in terms of the effective single-site model. Since the constant in \eqref{eq:single_Seff} depends on $\lambda$ and $g$, their contributions must also be retained. This leads to 
\begin{equation}
\overline{\text{SFF}}(T)=\text{Max}_{\lambda,g}~\left[\text{Tr}(e^{H^{(1)}_\text{eff}(\lambda)T})\right]^N\exp\left(\frac{N\Gamma_0T}{q}(g^q-1)-N\lambda T g\right)
\end{equation}
Here, the effective Hamiltonian $H^{(1)}_\text{eff}(\lambda)$ is given by replacing $\Gamma_0$ in \eqref{eq:partitionfunc_single_replica} with $\lambda$:
\begin{equation}
H^{(1)}_{\text{eff}}(\lambda)=i\sum_a\frac{\lambda}{2}\chi_{a}^L\chi_{a}^R-iU\chi_{1}^L\chi_{2}^L\chi_{3}^L\chi_{4}^L+iU\chi_{1}^R\chi_{2}^R\chi_{3}^R\chi_{4}^R.
\end{equation}
Note that the periodic boundary condition in \eqref{eq:defsff} translates into the trace over the single-site problem. Taking the saddle point of $\Sigma_a^{+-}$ and $G_a^{+-}$ then corresponds to maximizing the action with respect to $\lambda$ and $g$. Using the exact diagonalization, we could compute the trace exactly, which leads to 
\begin{equation}
\begin{aligned}
&e^{S_0(\lambda)}=\text{Tr}(e^{H^{(1)}_\text{eff}(\lambda)T})= \left(8 \cosh \left(\frac{1}{2} T \sqrt{ \lambda ^2-U^2}\right)+2\cosh (
   \lambda  T)+6\right),\\
&\ln \overline{\text{SFF}}(T)/N=\text{Max}_{\lambda,g}~\left[S_0(\lambda)+\frac{\Gamma_0T}{q}(g^q-1)-\lambda T g\right].
\end{aligned}
\end{equation}

The results obtained from direct numerical maximization are shown in Figure \ref{fig:SFF} for $q = 4$ and various values of $U / \Gamma_0$. In the short-time limit, the SFF is dominated by the diagonal saddle, where the forward and backward branches remain uncorrelated. This corresponds to $g = \lambda = 0$, which yields
\begin{equation}\label{eq:diag}
\left.\ln \overline{\text{SFF}}(T)/N\right|_{\text{diag.}}= \ln\left(8+8\cos\left(\frac{UT}{2}\right)\right)-\frac{\Gamma_0T}{q}.
\end{equation}
For $U = 0$, the diagonal contribution decays monotonically with increasing time, eventually giving way to the connected saddle with $\lambda, g \neq 0$ at $\Gamma_0 T \approx O(1)$. However, in the presence of a large $U$, the diagonal SFF exhibits persistent oscillations with multiple zeros located at $U T_n^* = 2(2n + 1)\pi$. Near each $T_n^*$, the diagonal saddle becomes subleading, allowing the connected saddle to dominate even at relatively small values of $\Gamma_0 T$. The solution for the connected saddle cannot be obtained analytically. Nevertheless, we can make an estimation by considering the long-time limit $T\rightarrow \infty$. This leads to $S_0(\lambda)\approx \lambda T$ and thus
\begin{equation}\label{eq:conn}
\left.\ln \overline{\text{SFF}}(T)/N\right|_{\text{conn.}}\approx\text{Max}_{\lambda,g}~\left[\frac{\Gamma_0T}{q}(g^q-1)+\lambda T (1-g)\right]=0,
\end{equation}
which corresponds to the saddle-point solution with $g=1$. From the numerical results, we observe that corrections to this value remain moderate even for $\Gamma_0 T \sim O(1)$.

We can further estimate the number of dynamical transitions by comparing the local maxima of the diagonal solution \eqref{eq:diag} with those of the connected contribution \eqref{eq:conn}. Guided by the numerical results, the locations of these maxima are found near $U \tilde{T}_n^* = 4n\pi$, with the maximal value approximately given by $4\ln 2 - \frac{4n\pi \Gamma_0}{Uq}$. Consequently, the condition for observing two dynamical transitions surrounding each $\tilde{T}_n^*$ requires $U / \Gamma_0 > n\pi / (q \ln 2)\approx 1.13\times n$ for $q=4$. Therefore, we expect the SFF for $U / \Gamma_0 = 1, 2, 3$ to exhibit one/three/five dynamical transitions between the two saddles, in agreement with the numerical observations shown in Figure \ref{fig:SFF}.

\section{Out-of-time-order Correlator}
The out-of-time-order correlator (OTOC) has garnered extensive attention across multiple disciplines for its capacity to quantify information scrambling, probe near-horizon dynamics of black holes, and connect to diverse information-theoretic tasks \cite{LaOv69,kitaev2015hidden,Shenker:2013pqa,Shenker:2014cwa,Hayden:2007cs,Hosur:2015ylk,Yoshida:2017non,Gao:2016bin,Maldacena:2017axo}. In systems with a large local Hilbert space dimension, the OTOC exhibits a universal regime of exponential deviation, characterized by the quantum Lyapunov exponent $\varkappa$. Ref.~\cite{Maldacena:2015waa} established an upper bound on this exponent, $\varkappa \leq 2\pi/\beta$, which is saturated by holographic systems. Consequently, the quantum Lyapunov exponent serves as a key diagnostic for identifying systems with holographic duals. Subsequently, Ref.~\cite{Zhang:2020jhn} argued that the realization of sub-AdS holography further requires a large branching time $t_B$, whose definition will be introduced below. However, it has been shown that SYK-like models with simple ladder-diagram structures in the OTOC satisfy a bound $t_B (\varkappa+\Gamma) \leq 2$. This motivates the study of OTOCs beyond ladder diagrams, and a concrete example, as we elaborate below, is provided by the Brownian SYK–Hubbard model.

We focus on the retarded out-of-time-order correlator (OTOC) for single Majorana operators, defined as \cite{Kitaev:2017awl}
\begin{equation}
\text{OTOC}_{ab}(t)=\frac{\Theta(t)}{N}\sum_{ij}\left<\{\chi_{ia}(t),\chi_{jb}(0)\}\{\chi_{ia}(t),\chi_{jb}(0)\}\right>,
\end{equation}
where $\Theta(t)$ denotes the Heaviside step function. Owing to the permutation symmetry, the result depends only on whether $a=b$ or not. The factor of $1/N$ is introduced to gurantee a finite result in the large-$N$ limit. The OTOC involves two forward and two backward time evolutions. Consequently, its path-integral representation requires introducing a double Keldysh contour \cite{Aleiner:2016eni}. We start by analyzing the partition function $1=\overline{Z}^{(2)}=\overline{\text{tr}[U(t)U^\dagger(t)U(t)U^\dagger(t)]}$. After introducing the auxiliary fields, the partition function can be expressed as
\begin{equation}\label{eq:Ztworeplica_pathintegral}
\begin{aligned}
\overline{Z}^{(2)}&=\ \begin{tikzpicture}[scale=1.3, baseline={([yshift=-2pt]current bounding box.center)}]
\draw[thick] (20pt,-5pt) arc(90:-90:5pt and 5pt);
\draw[thick] (-20pt,5pt) arc(90:270:5pt and 5pt);
\draw[mid arrow,thick]  (-20pt,5pt)  --  (20pt,5pt);
\draw[mid arrow,thick]  (20pt,-5pt)  --  (-20pt,-5pt);

\draw[thick] (20pt,-25pt) arc(-90:90:15pt and 15pt);
\draw[thick] (-20pt,-15pt) arc(90:270:5pt and 5pt);
\draw[mid arrow,thick]  (-20pt,-15pt)  --  (20pt,-15pt);
\draw[mid arrow,thick]  (20pt,-25pt)  --  (-20pt,-25pt);

\draw[thick] (-25pt,5pt) node[left]{\scriptsize$0$};
\draw[thick] (30pt,5pt) node[right]{\scriptsize$t$};

\draw[thick] (-25pt,-25pt) node[left]{};
\filldraw (35pt,-10pt) circle (1pt);
\end{tikzpicture}~=\int  \mathcal{D}\chi_{ia}^s  \mathcal{D}\Sigma_a^{ss'}\mathcal{D}G_a^{ss'}~\exp\left(-S^{(2)}\right),\\
S^{(2)}&=\int dt~\left[\frac{1}{2}\chi^s_{ia}(f_s\delta_{ss'}\partial_t-\Sigma_{a}^{ss'})\chi^{s'}_{ia}+\frac{1}{2}\Sigma_{a}^{ss'}G_{a}^{ss'}+\frac{Jf_sf_{s'}}{2q}(G_{a}^{ss'})^q+iUf_s\chi^s_{i1}\chi^s_{i2}\chi^s_{i3}\chi^s_{i4}\right].
\end{aligned}
\end{equation}
Here, $s \in \{1,2,3,4\}$ labels the four branches of the contour, where $s=1,3$ correspond to forward evolutions with $f_s=1$ and $s=2,4$ to backward evolutions with $f_s=-1$. Similar to the calculation on the single Keldysh contour, the saddle-point solution can be evaluated explicitly owing to the connectivity between the forward and backward evolutions. Consequently, we obtain
\begin{equation}
\Sigma_{a}^{ss'}=-\Sigma_{a}^{ss'}=\frac{\Gamma_0}{2},\ \ \ \ \ \ G_{a}^{ss'}=-G_{a}^{ss'}=\frac{1}{2},
\end{equation}
for $s<s'$, with the sign determined by the contour ordering. 

The single-site evolution is then local in time and can again be described by an effective Hamiltonian, now defined on sixteen modes with $a, s \in \{1, 2, 3, 4\}$. This evolution can also be interpreted as a generalized Lindblad equation for multiple replicas \cite{Zhou:2022qhe}. Focusing on a representative site $j_0$, we identify the Majorana fields $(\chi_{j_0a}^1, \chi_{j_0a}^2, \chi_{j_0a}^3, \chi_{j_0a}^4)$ with the Majorana operators $(\chi_{a}^{L_1}, i\chi_{a}^{R_1}, \chi_{a}^{L_2}, i\chi_{a}^{R_2})$, and map the partition function $\overline{Z}^{(2)}$ to the inner product:
\begin{equation}\label{eq:Ztworeplica}
\begin{aligned}
\tilde Z&=\ \begin{tikzpicture}[scale=1.3, baseline={([yshift=-2pt]current bounding box.center)}]
\draw[thick] (20pt,-5pt) arc(90:-90:5pt and 5pt);
\draw[thick] (-20pt,5pt) arc(90:270:5pt and 5pt);
\draw[mid arrow,thick]  (-20pt,5pt)  --  (20pt,5pt);
\draw[mid arrow,thick]  (-20pt,-5pt)  --  (20pt,-5pt);

\draw[thick] (20pt,-25pt) arc(-90:90:15pt and 15pt);
\draw[thick] (-20pt,-15pt) arc(90:270:5pt and 5pt);
\draw[mid arrow,thick]  (-20pt,-15pt)  --  (20pt,-15pt);
\draw[mid arrow,thick]  (-20pt,-25pt)  --  (20pt,-25pt);

\draw[dotted,thick] (-18pt,7pt) -- (-18pt,-27pt);
\draw[dotted,thick] (20pt,7pt) -- (20pt,-27pt);
\end{tikzpicture}~\propto~\bra{\text{EPR}_2}e^{H_{\text{eff}}^{(2)}t}\ket{\text{EPR}_1}^N.
\end{aligned}
\end{equation}
Here, the effective Hamiltonian $H_{\text{eff}}^{(2)}$ contains full coupling between four branches due to the self-energy term $\Sigma_{a}^{ss'}$:
\begin{equation}\label{eq:H2eff}
\begin{aligned}
H_{\text{eff}}^{(2)}=&-\frac{\Gamma_0}{2}\sum_a(\chi_{a}^{L_1}-i\chi_{a}^{R_1})(\chi_{a}^{L_2}-i\chi_{a}^{R_2})+i\frac{\Gamma_0}{2}\sum_a(\chi^{L_1}_a\chi^{R_1}_a+\chi^{L_2}_a\chi^{R_2}_a)
\\&-iU\chi_{1}^{L_1}\chi_{2}^{L_1}\chi_{3}^{L_1}\chi_{4}^{L_1}+iU\chi_{1}^{R_1}\chi_{2}^{R_1}\chi_{3}^{R_1}\chi_{4}^{R_1}-iU\chi_{1}^{L_2}\chi_{2}^{L_2}\chi_{3}^{L_2}\chi_{4}^{L_2}+iU\chi_{1}^{R_2}\chi_{2}^{R_2}\chi_{3}^{R_2}\chi_{4}^{R_2}.
\end{aligned}
\end{equation} 
The couplings between the two forward (or two backward) evolution branches are anti-Hermitian, whereas those between forward and backward branches are Hermitian. Different boundary conditions at times $0$ and $t$ in the path-integral representation \eqref{eq:Ztworeplica_pathintegral} lead to distinct initial and final EPR states in \eqref{eq:Ztworeplica}. The initial state $\ket{\text{EPR}_1}$ is defined as the EPR pairing between $(L_1, R_1)$ and $(L_2, R_2)$:
\begin{equation}
(\chi_{a}^{L_1}-i\chi_{a}^{R_1})\ket{\text{EPR}_1}=0,\ \ \ \ \ \ (\chi_{a}^{L_2}-i\chi_{a}^{R_2})\ket{\text{EPR}_1}=0,
\end{equation}
while the final state $\ket{\text{EPR}_2}$ is defined as the EPR pairing between $( R_1,L_2)$ and $(L_1, R_2)$:
\begin{equation}
(\chi_{a}^{L_1}-i\chi_{a}^{R_2})\ket{\text{EPR}_2}=0,\ \ \ \ \ \ (\chi_{a}^{R_1}-i\chi_{a}^{L_2})\ket{\text{EPR}_2}=0.
\end{equation}
From \eqref{eq:H2eff}, it is straightforward to verify that $\ket{\text{EPR}_1}$ is a right eigenstate of the effective Hamiltonian $H_{\text{eff}}^{(2)}$. We can further rearrange $H_{\text{eff}}^{(2)}$ into 
\begin{equation}
\begin{aligned}
H_{\text{eff}}^{(2)}=&i\frac{\Gamma_0}{2} \sum_a(\chi_{a}^{L_1}+i\chi_{a}^{R_2})(\chi_{a}^{R_1}+i\chi_{a}^{L_2})+i\frac{\Gamma_0}{2}\sum_a(\chi^{L_1}_a\chi^{R_2}_a+\chi^{R_1}_a\chi^{L_2}_a)
\\&-iU\chi_{1}^{L_1}\chi_{2}^{L_1}\chi_{3}^{L_1}\chi_{4}^{L_1}+iU\chi_{1}^{R_1}\chi_{2}^{R_1}\chi_{3}^{R_1}\chi_{4}^{R_1}-iU\chi_{1}^{L_2}\chi_{2}^{L_2}\chi_{3}^{L_2}\chi_{4}^{L_2}+iU\chi_{1}^{R_2}\chi_{2}^{R_2}\chi_{3}^{R_2}\chi_{4}^{R_2}.
\end{aligned}
\end{equation} 
This demonstrates that $\ket{\text{EPR}_2}$ is a left eigenstate of the effective Hamiltonian $H_{\text{eff}}^{(2)}$.

  \begin{figure}[t]
    \centering
    \includegraphics[width=0.83\linewidth]{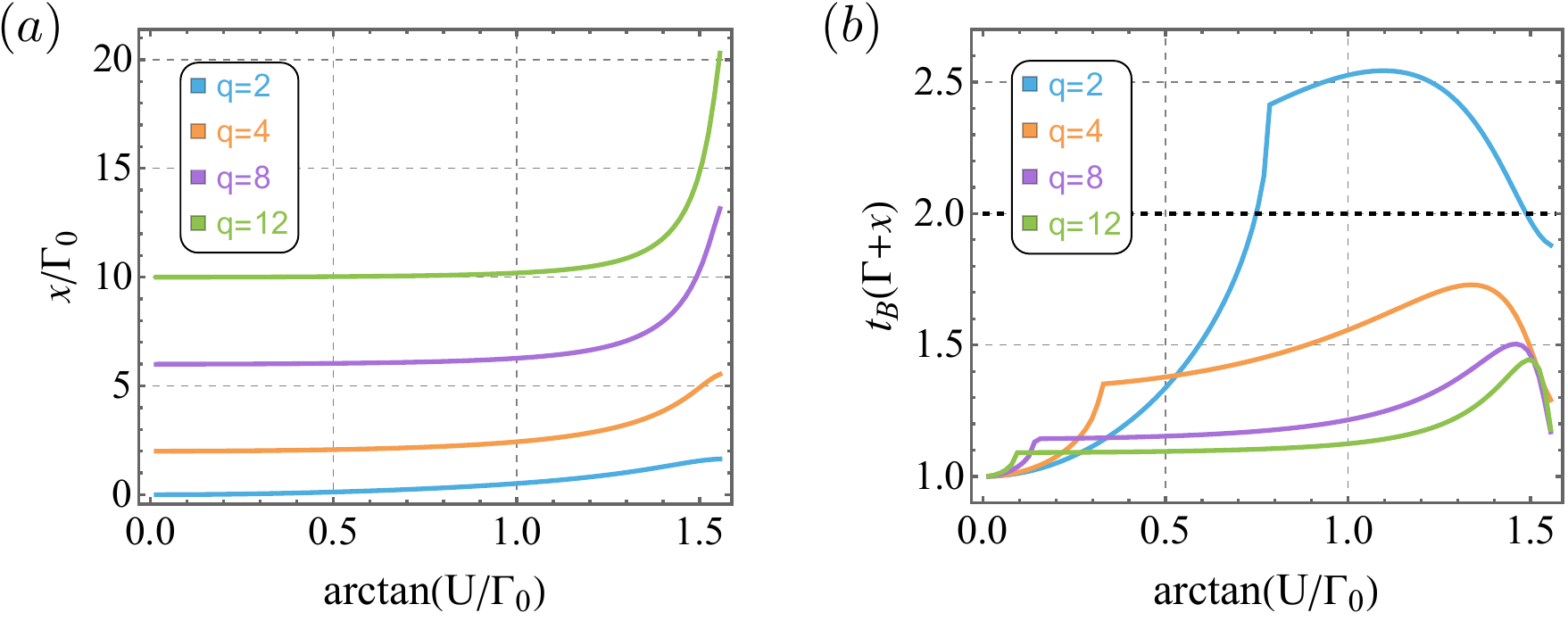}
    \caption{ We plot the quantum Lyapunov exponent $\varkappa$ and the branching time $t_B$ as functions of $U/\Gamma_0$ for the Brownian SYK–Hubbard model with $q \in {2,4,8,12}$. The results reveal that increasing the Hubbard interaction $U$ enhances many-body chaos and, notably, leads to a violation of the branching-time bound for $q=2$, as indicated by the black dashed line.}
    \label{fig:otoc}
  \end{figure}

Next, we compute the OTOC for the Brownian SYK–Hubbard model. In contrast to the two-point function, the full OTOC receives substantial contributions from fluctuations beyond the saddle-point solution \cite{Maldacena:2016hyu}, and thus cannot be expressed solely by inserting additional operators into the inner product \eqref{eq:Ztworeplica}. We begin with a diagrammatic analysis of the OTOC. By expanding the OTOC in terms of the Brownian SYK couplings, we can write out the self-consistent equation for a series of generalized ladder diagrams: 
\begin{equation}\label{eq:consistentOTOC}
\begin{aligned}
\begin{tikzpicture}
\node[dvertex] (R) at (-30pt,0pt) {\scriptsize $\ $};
\draw[thick] (R) -- ++(135:20pt) node[left]{\scriptsize$t,ia$};
\draw[thick] (R) -- ++(-135:20pt) node[left]{\scriptsize$t,ia$};
\draw[thick] (R) -- ++(45:20pt) node[right]{\scriptsize$0,jb$};
\draw[thick] (R) -- ++(-45:20pt) node[right]{\scriptsize$0,jb$};
\end{tikzpicture}&=\delta_{ij}\begin{tikzpicture}
\node[bvertex] (R) at (-30pt,0pt) {\scriptsize$F_{ab}$};
\draw[thick] (R) -- ++(135:20pt) node[left]{\scriptsize$t,ia$};
\draw[thick] (R) -- ++(-135:20pt) node[left]{\scriptsize$t,ia$};
\draw[thick] (R) -- ++(45:20pt) node[right]{\scriptsize$0,ib$};
\draw[thick] (R) -- ++(-45:20pt) node[right]{\scriptsize$0,ib$};
\end{tikzpicture}+
\begin{tikzpicture}
\node[bvertex] (R) at (-30pt,0pt) {\scriptsize$F_{ac}$};
\node[dvertex] (A) at (30pt,0pt) {\scriptsize $\ $};
\draw[thick] (R) -- ++(135:20pt) node[left]{\scriptsize$t,ia$};
\draw[thick] (R) -- ++(-135:20pt) node[left]{\scriptsize$t,ia$};
\draw[thick] (A) -- ++(45:20pt) node[right]{\scriptsize$0,jb$};
\draw[thick] (A) -- ++(-45:20pt) node[right]{\scriptsize$0,jb$};
\draw[thick] (A) to[out=140,in=40] (R);
\draw[thick] (A) to[out=-140,in=-40] (R);
\draw[wavy] (0,15pt) -- (0,-15pt);
\node at (-13pt,20pt) {\scriptsize$t',ic$};
\node at (-13pt,-20pt) {\scriptsize$t',ic$};
\node at (13pt,20pt) {\scriptsize$t',kc$};
\node at (13pt,-20pt) {\scriptsize$t',kc$};
\end{tikzpicture}
,\\
\text{OTOC}_{ab}(t)&=F_{ab}(t)+(q-1)\Gamma_0\sum_{c}\int_{0}^t dt'~F_{ac}(t-t')\text{OTOC}_{cb}(t').
\end{aligned}
\end{equation}
Here, $F_{ab}(t)$ denotes the single-site OTOC, evaluated to leading order in the $1/N$ expansion:
\begin{equation}
\begin{aligned}
F_{ab}(t)&=\Theta(t)\left<\{\chi_{j_0a}(t),\chi_{j_0b}(0)\}\{\chi_{j_0a}(t),\chi_{j_0b}(0)\}\right>.
\end{aligned}
\end{equation}
If the Hubbard interaction is turned off, the single-site problem becomes non-interacting and can be computed using Wick’s theorem. This yields $F_{ab}(t)=-\delta_{ab}G^R(t)^2=\delta_{ab}\Theta(t)e^{-\Gamma_0t}$, where the retarded Green’s function is $G^R(t)=-i\Theta(t)\left<\{\chi_{j_0a}(t),\chi_{j_0a}(0)\}\right>=-i\Theta(t)e^{-\frac{\Gamma_0t}{2}}$. This result corresponds precisely to the “rung” of the ladder diagrams in standard SYK calculations \cite{Maldacena:2016hyu,Zhang:2020jhn}. When the Hubbard interaction is turned on, we can instead express $F_{ab}(t)$ using the effective Hamiltonian $H_{\text{eff}}^{(2)}$, which leads to 
\begin{equation}\label{eqn:Fab}
\begin{aligned}
F_{ab}(t)=&\Theta(t)\frac{\bra{\text{EPR}_2}(\chi_{a}^{L_1}-i\chi_{a}^{R_1})(\chi_{a}^{L_2}-i\chi_{a}^{R_2})e^{H_{\text{eff}}^{(2)}t}\chi_{b}^{L_2}\chi_{b}^{L_1}\ket{\text{EPR}_1}}{\bra{\text{EPR}_2}e^{H_{\text{eff}}^{(2)}t}\ket{\text{EPR}_1}}\\=&\Theta(t)\frac{\bra{\text{EPR}_2}(\chi_{a}^{L_1}-i\chi_{a}^{R_1})(\chi_{a}^{L_2}-i\chi_{a}^{R_2})e^{(H_{\text{eff}}^{(2)}-2\Gamma_0)t}\chi_{b}^{L_2}\chi_{b}^{L_1}\ket{\text{EPR}_1}}{\braket{\text{EPR}_2|\text{EPR}_1}}.
\end{aligned}
\end{equation}

We are interested in the exponential growth regime of the OTOC at relatively long times. In this regime, we expect $\text{OTOC}(t)\equiv \sum_b \text{OTOC}_{ab}(t)\sim C_0 e^{\varkappa t}$, where $C_0$ is a constant independent of $a$, owing to the permutation symmetry among the four Majorana modes. When the exponential term becomes sufficiently large, the inhomogeneous contribution $F_{ab}(t)$ in the self-consistent equation \eqref{eq:consistentOTOC} can be neglected. This leads to the homogenuous equation:
\begin{equation}\label{eq:homogeneous}
\text{OTOC}(t)=(q-1)\Gamma_0\int_{0}^t dt'~F(t-t')\text{OTOC}(t').
\end{equation}
where, for conciseness, we define $F(t)=\sum_{c=1}^4F_{1c}(t)$. Consequently, the OTOC becomes an eigenfunction of the integral operator $(q-1)\Gamma_0F(t-t')$ with eigenvalue $1$. Owing to time-translation invariance, the eigenfunctions of this operator take the form of simple exponential functions:
\begin{equation}
k_R(h)f_h(t)=(q-1)\Gamma_0\int_{0}^t dt'~F(t-t')f_h(t'),\ \ \ \ \ \ f_h(t)=e^{-h t},
\end{equation}
where the corresponding eigenvalue $k_R(h)$ can be obtained by substituting \eqref{eqn:Fab} into the integrand. The resulting expression is given by
\begin{equation}
\begin{aligned}
k_R(h)&=\Gamma_0(q-1)\frac{\bra{\text{EPR}_2}(\chi_{a}^{L_1}-i\chi_{a}^{R_1})(\chi_{a}^{L_2}-i\chi_{a}^{R_2})(2\Gamma_0-h-H_{\text{eff}}^{(2)})^{-1}\chi_{b}^{L_2}\chi_{b}^{L_1}\ket{\text{EPR}_1}}{\braket{\text{EPR}_2|\text{EPR}_1}}\\
&=\frac{\Gamma _0 (q-1) \left(\left(h-2 \Gamma _0\right){}^2 \left(3 \Gamma _0-h\right)+U^2
   \left(7 \Gamma _0-2 h\right)\right)}{\left(h-2 \Gamma _0\right){}^2 \left(3 \Gamma
   _0^2+h^2-4 \Gamma _0 h\right)+U^2 \left(\Gamma _0^2+h^2-4 \Gamma _0 h\right)}.
\end{aligned}
\end{equation}
Here, the result in the second line is obtained from exact diagonalization. The quantum Lyapunov exponent is determined by solving $k_R(-\varkappa) = 1$, while the branching time, which characterizes the stability of the quantum Lyapunov exponent, is defined as $t_B = k_R'(-\varkappa)$.

The numerical results for the quantum Lyapunov exponent and the branching time are shown in Figure \ref{fig:otoc} for several values of $q$. In the limit $U \to 0$, we recover $\varkappa = (q - 2)\Gamma_0$ and $t_B (\varkappa + \Gamma_0) = 1$, consistent with known results in the literature \cite{Zhang:2020jhn}. Upon introducing the Hubbard interaction, $\varkappa$ increases monotonically, in line with the physical intuition that stronger interactions enhance quantum many-body chaos. Notably, for $q = 2$, the Brownian SYK coupling acts as a hopping term, and chaos emerges solely from the Hubbard interaction. The branching time, multiplied by $\varkappa + \Gamma_0$, exhibits a non-analyticity inherited from the non-analytic behavior of $\Gamma$, as discussed in the analysis of the two-point function. Furthermore, it displays a peak at intermediate $U/\Gamma_0$, which can violate the bound $t_B (\varkappa + \Gamma_0) \leq 2$ derived using ladder diagrams with simple rung functions. Because this bound is satisfied in traditional SYK-like models, the observed violation clearly demonstrates that the Brownian SYK–Hubbard model constitutes a new class of models beyond the original SYK paradigm.

\section{Discussions}
In this work, we introduce the Brownian SYK–Hubbard model as a solvable framework that (1) enables the study of Hubbard interactions in chaotic many-body systems and (2) extends beyond the conventional SYK large-$N$ structure. In particular, we compute the two-point function, the SFF, and the OTOC by mapping the path integral to an effective Hamiltonian with a few sites, which allows for exact diagonalization. The two-point function in the time domain undergoes a dynamical transition from monotonic decay to oscillatory behavior as the Hubbard interaction increases, corresponding to a change from a single-peak to a double-peak structure in the single-particle spectral function. The SFF exhibits a series of first-order transitions between the diagonal and connected saddles, with the number of transitions depending on the Hubbard interaction $U$. For the OTOC, we observe an enhancement of many-body chaos as $U$ increases and provide an explicit example of the violation of the bound on branching time. These results establish a new analytically tractable platform for exploring the interplay of Hubbard interactions and chaos in many-body systems.

We conclude with several remarks. First, in this work we have combined the Hubbard interaction with Brownian SYK couplings, rather than with static SYK interactions. In the latter case, the self-energy field becomes bilocal in time and cannot be solved analytically. This situation closely resembles that encountered in dynamical mean-field theory \cite{RevModPhys.68.13}, which typically requires numerical approaches such as quantum Monte Carlo methods. Second, while our analysis has focused on correlation functions and the spectral form factor, it would be interesting to explore the entanglement dynamics by generalizing the methods developed in Ref.~\cite{Jian:2020krd}. In particular, investigating the fate of replica wormhole–like solutions \cite{Chen:2020wiq} as the Hubbard interaction increases is a compelling direction for future work. Finally, our approach-combining a small system solvable by exact diagonalization with SYK-like random interactions among such subsystems-provides a general framework for constructing analytically tractable models. Extending this strategy to systems directly related to superconductivity or magnetism would be especially intriguing and is left for future investigation.

\section*{Acknowledgment}
We thank Yingfei Gu and Tian-Gang Zhou for helpful discussions. This project is supported by the NSFC under grant 12374477, the Shanghai Rising-Star Program under grant number 24QA2700300, the Innovation Program for Quantum Science and Technology 2024ZD0300101, the Shanghai Municipal Science and Technology Major Project Grant No. 24DP2600100, the National Key Research and Development Program of China Grant No. 2022YFA1405300, and NSAF Grant No. U1930201.

\bibliography{Draft.bbl}

\end{document}